\documentclass[a4paper,11pt]{article}
\usepackage[utf8]{inputenc}
\usepackage{graphicx,xcolor}
\usepackage{setspace}
\usepackage{amssymb}
\usepackage{amsmath}
\usepackage[round]{natbib}
 
\date{}

\title{On the influence of cell size on network modelling of water retention and conductivity of geomaterials} 

\author{Ignatios Athanasiadis, Simon Wheeler, Peter Grassl$^{*}$}

\begin{document}
\maketitle
\begin{center}
School of Engineering, University of Glasgow, Glasgow, UK\\
$^{*}$ Corresponding author, Email: peter.grassl@glasgow.ac.uk
\end{center}

\section*{Abstract}
Network modelling of water retention and conductivity of porous geomaterials based on filling and emptying of pores requires the use of a computational cell which is a representative volume element of the material. The conditions for the existence of a representative volume element are that with increasing cell size the statistical results of multiple random cells approach a constant mean and zero coefficient of variation of the macroscopic property. In this study, random network analyses were carried out to determine the mean and coefficient of variation of degree of saturation and conductivity for varying capillary suction and cell size.
For water retention, a representative volume element was obtained for the entire range of capillary suction investigated.
For conductivity, a representative volume element was obtained for most of the range of capillary suction, but there is a small range of capillary suction for which the coefficient of variation does not reduce with increasing size of the computational cell, due to onset and termination of percolation of the wetting fluid (during wetting and drying, respectively).

Keywords: Network model; representative volume element; porous geomaterial; transport; unsaturated; periodic boundary conditions

\section{Introduction}
The response of porous materials is strongly governed by processes occurring at the pore-scale.
For unsaturated conditions, transport properties are determined by filling and emptying of individual pores, which results in hysteresis of fluid retention and hydraulic conductivity \citep{Lev41,Pou62}.
Transport network models have been widely used to predict these phenomena \citep{Fat56P1,Fat56P2,Fat56P3,TsaPay90,JerSal90,BluKin91,BryBlu92}. Previous authors have investigated the influence of network properties such as mean coordination number, and sphere and pipe size distributions and spatial correlation on retention and relative permeability curves, including the influence of phenomena such as trapping and snap off, as discussed in \citet{Blu01,Blu02}. With the appropriate input, these network models are capable of predicting the influence of processes at porescale on macroscopic properties.

In the network model used in this study, the microstructure of a porous geomaterial is idealised by a computational cell with statistical distributions of sphere radii, representing the pores, and pipe radii, representing the narrower throats forming the connections between pores.
Each random cell results in a different microstructure, because the generation of sphere locations and sphere and pipe radii is random.
For a series of network analyses, all with the same size of cell, statistics in the form of the mean and the coefficient of variation (COV) can be evaluated for a property of interest, such as degree of saturation $S_{\rm r}$ or conductivity $k_{\rm {CELL}}$. If a representative volume element of the material exists, the mean approaches a constant and the COV approaches zero for increasing cell size.  

For saturated conditions, it is known that network models provide a representative volume element \citep{ZhaZhaChe00,DuOst06} for conductivity, because as the cell size increases, more information of the chosen distributions is introduced and the differences of the microstructure between cells become less significant.
In the limit of an infinitely large cell, the COV of the quantity would be zero, which corresponds to the deterministic representative volume element (RVE), for which a single cell of that size can adequately represent the chosen microstructure. For finite cell sizes, the non-zero COV can be used to define a statistical representative volume element (SRVE), i.e. a cell size at which the COV is smaller than a prescribed value.

For unstaturated conditions, less work has been carried out to investigate the influence of cell size on water retention and conductivity. The aim of this study is to investigate the influence of cell size on degree of saturation and conductivity for varying capillary suction representing varying unsaturated conditions. The response will be predicted with a network model based on three-dimensional Voronoi tessellations developed in \cite{AthWheGra16,Ath17,AthWheGra18}.

\section{Model} \label{sec:model}
The retention and conductivity of cells of porous geomaterial are modelled by a network of pipes and spheres with statistical distribution of radii \citep{AthWheGra16,Ath17}.
The network is generated by a Voronoi tessellation of randomly placed points which are subject to a minimum distance $d_{\rm min}$.
The edges of Voronoi polyhedra are used as the locations for pipes and the vertices of the polyhedra as the locations for spheres.
Statistical log-normal distributions of radii for spheres and pipes with mean $r_{\rm m}$ and coefficient of variation COV were used subject to lower and upper radii cut-offs $r^{\rm min}$~and~$r^{\rm max}$, respectively.
Sphere radii were allocated randomly to individual locations within the tessellation.
Pipe radii were then allocated to individual locations using a procedure described by \cite{Ath17}, which was intended to minimize any risk of a pipe of relatively large radius connecting to a sphere of smaller radius.

The network was used to determine the degree of saturation and conductivity of a computational cell, which was subjected to monotonically decreasing or increasing capillary suction $P_{\rm c}$ applied uniformly across the cell for wetting or drying, respectively.
The capillary suction $P_{\rm c}$ is defined as $P_{\rm c} = P_{\rm a}-P_{\rm w}$, where $P_{\rm a}$ is the pressure in the drying fluid (air) and $P_{\rm w}$ is the pressure in the wetting fluid (water). Here, $P_{\rm a}$ is assumed to be equal to the atmospheric pressure. The value of pressures is assumed to be positive so that a positive value of $P_{\rm c}$ corresponds to a negative water pressure, i.e. tension. 

At the scale of a boundary value problem, gradients of capillary suction are commonly present, which are the driving force for wetting and drying processes.
However, in this study, when assessing the retention behaviour of the cells, these gradients are not considered at the level of the cell.
Instead, a uniform capillary suction applied to the entire cell is used to control the wetting and drying process and to determine the degree of saturation.
The conductivity of the cell is subsequently determined by separately applying a numerical water pressure gradient across the cell.

With the present model, the cell size is much smaller than a realistic boundary value problem. Therefore, it was decided to use a periodic cell with a periodic material structure and periodic boundary conditions (PBC). A recently proposed approach to apply PBC was used, which was originally presented in \citep{AthWheGra16,AthWheGra18}. Here, elements are allowed to cross the boundaries of the cell. Each element that crosses a boundary has at least one other element that it is periodic to (has the same length and orientation) that crosses the cell boundary at another point located diagonally or longitudinally across the cell.  With this technique, networks can maintain their random geometry while having a periodic structure.

With the use of PBC, there are no external boundaries from which to introduce the intruding fluid. Instead, a drying path must start with at least one sphere within the cell already filled with the drying fluid and a wetting path must start with at least one sphere already filled with the wetting fluid.
In \cite{Ath17} it was shown that realistic modelling of a drying or wetting path could not be achieved by starting with only a single sphere filled with the intruding fluid, because of the extremely low connectivity (only 4 pipes) from a single sphere and the consequent possibility that extreme changes of capillary suction $P_{\rm c}$ might be required for the intruding fluid to break out from this first sphere (dependent on only the radii of the 4 pipes in the case of a drying path or the radii of the 4 spheres on the other ends of these pipes in the case of a wetting path) or to break out from the local region immediately surrounding the first sphere.
In order to achieve realistic modelling, drying or wetting paths should start with a small number of spheres already filled with the intruding fluid and these “seeding” spheres should not simply be selected at random. Instead, to achieve realistic modelling of a drying path, it should always be preceded by modelling of a previous wetting path finishing with a small number of spheres (the seeding spheres) still filled with the drying fluid.
A similar condition applies for realistic modelling of a wetting path, which should always be preceded by modelling of a previous drying path to leave an appropriate number of seeding spheres filled with the wetting fluid. Based on investigations presented in \cite{Ath17}, the volume of seeding spheres was selected as 1\% of the total void volume.

For wetting, the specimen was initially almost dry, i.e. only 1\% of the volume of spheres was filled with wetting fluid. It was then subjected to a monotonically decreasing capillary suction $P_{\rm c}$. For drying, the specimen was initially almost wet, i.e. only 1\% of the volume of spheres was filled with drying fluid and it was then subjected to a monotonically decreasing capillary suction $P_{\rm c}$.
At different values of $P_{\rm c}$ during the wetting and drying process, the transport network was subjected to a numerical wetting fluid pressure gradient to determine the conductivity of the cell. 
In the following sections, the individual parts of the transport network are described.

\subsection{Retention model}\label{sec:retentionModel}
Capillary suction $P_{\rm c}$, applied uniformly throughout the cell, was the driving variable for wetting and drying of the cell.
The rules that determine, for a given value of $P_{\rm c}$, which spheres are filled with the wetting and non-wetting fluid, respectively, are based on the Young-Laplace equation \citep{Was21}
\begin{equation}\label{eq:LY}
  P_{\rm c} = \dfrac{2 \gamma \cos \theta}{r}
\end{equation}
where $\gamma$ is the surface tension and $\theta$ is the contact angle formed between the fluid-fluid interface and the solid (measured on the wetting fluid side).
Furthermore, $r$ is the radius of a pipe during a drying process and the radius of a sphere during a wetting process, to represent the so-called “ink bottle” effect \citep{AbeWilLan99}.
A drying or wetting process is modelled by considering the intruding fluid moving from a void already filled with that fluid to a connected void not previously filled with that fluid, i.e. direct connection to a void already filled with the intruding fluid is imposed as a requirement for a void to fill with the intruding fluid.
The drying fluid is the intruding fluid during a drying path, whereas the wetting fluid is the intruding fluid during a wetting path.

\begin{figure}
\centering
  \begin{tabular}{cc}
  \includegraphics[width=5cm]{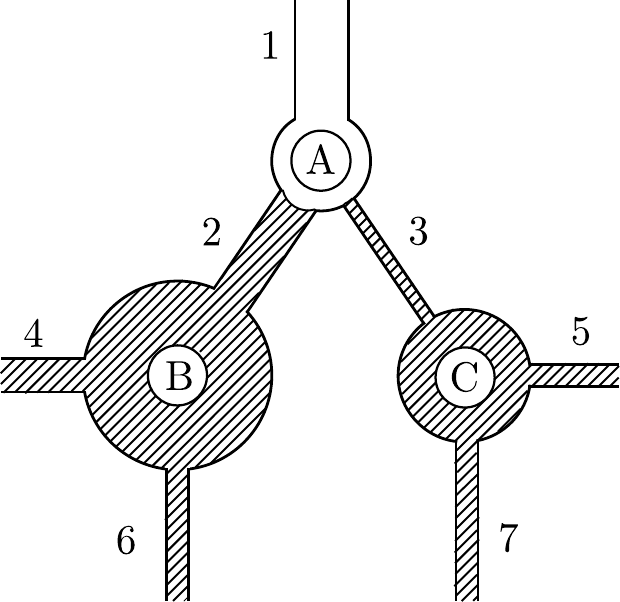} & \includegraphics[width=5cm]{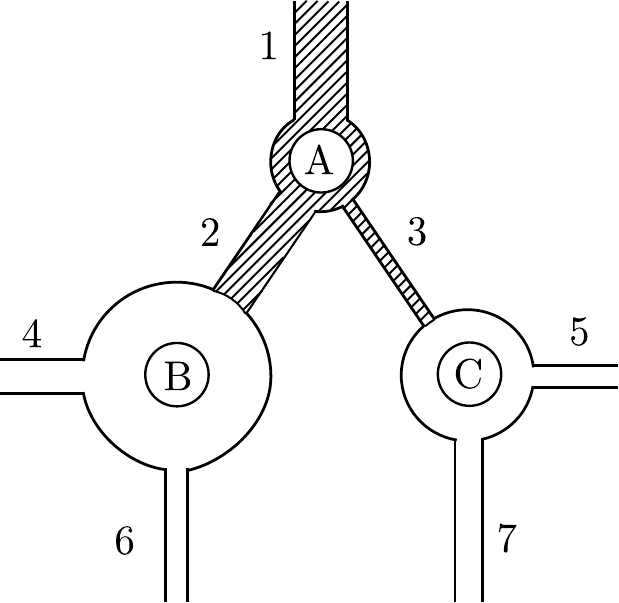}\\
  (a) & (b)
  \end{tabular}
  \caption{Retention behaviour: A simple 2-D network subjected to (a) drying and (b) wetting. The hatched areas represent the wetting fluid. The drying and wetting fluid for cases (a) and (b), respectively, is allowed to enter the network from pipe~1.}
  \label{fig:retention}
\end{figure}

Figure~\ref{fig:retention}a illustrates the situation during a drying process. The hatched area corresponds to the wetting fluid. The drying fluid has already intruded through pipe~1 into sphere~A at a value of $P_{\rm c}$ corresponding to (\ref{eq:LY}) with $r$ as the radius of pipe~1. As the value of $P_{\rm c}$ is increased further during the drying process, the next consideration is when the drying fluid will intrude further from sphere~A along either pipe~2 or pipe~3. Pipe~2 is of larger radius than pipe~3, and therefore the first action to occur is intrusion of the drying fluid along pipe~2 and into sphere~B, at a value of $P_{\rm c}$ corresponding to (\ref{eq:LY}) with $r$ as the radius of pipe~2. If either pipe~4 or pipe~6 was larger than pipe~2 the intrusion would continue along this pipe into an additional sphere without need for further increase of $P_{\rm c}$.

Figure~\ref{fig:retention}b illustrates an equivalent situation for a wetting process, with the hatched area again representing the wetting fluid. Wetting fluid has already entered sphere~A from pipe~1, at a value of $P_{\rm c}$ corresponding to (\ref{eq:LY}) with $r$ as the radius of sphere~A, and immediately moved into pipes~2~and~3. As the value of $P_{\rm c}$ is reduced further during the wetting process, the next consideration is when the wetting fluid will intrude into sphere~B or sphere~C. As sphere~C is smaller than sphere~B, the first thing that happens is intrusion of the wetting fluid into sphere~C, at a value of $P_{\rm c}$ corresponding to (\ref{eq:LY}) with $r$ as the radius of sphere~C, and immediate further movement of the wetting fluid into pipes~5~and~7. If either pipe~5 or pipe~7 connected to an additional sphere that was smaller than sphere~C, the wetting fluid would continue into this sphere without need for further decrease of $P_{\rm c}$.

At each value of $P_{\rm c}$ during a drying or wetting process, the degree of saturation $S_{\rm r}$ (expressed in terms of the wetting fluid) is calculated by considering the proportion of spheres filled with the wetting fluid.
Hence, if $m$ is the total number of spheres in the cell and $m_{\rm w}$ is the number of these spheres filled with the wetting fluid, the degree of saturation $S_{\rm r}$ is given by

\begin{equation} \label{eq:saturation}
S_{\rm r} = \dfrac{\sum\limits^{m_{\rm w}}_{i=1} V_{\rm{v}i}}{\sum\limits^{m}_{k=1} V_{\rm{vk}}}
\end{equation}
where $V_{\rm v}$ is the volume of the voids corresponding to sphere $i$.

In analysing drying and wetting processes, no possibility of trapping of the non-intruding fluid is considered, because there is no consideration of whether there is appropriate connectivity to provide an exit route for the non-intruding fluid (connectivity requirements are applied to the intruding fluid, but not to the non-intruding fluid). In practice, trapping of the non-intruding fluid is not an issue if the non-intruding fluid can escape by diffusing through the intruding fluid, as is the case for water and air, and provided that drying or wetting is performed sufficiently slowly for the diffusion process to dissipate excess pressure in the trapped non-intruding fluid.

For a material with a wide range of void sizes, a cell would require a very large number of smaller voids in order to also include a statistically representative number of larger voids. With a standard network, this would require a cell with an impractically large number of spheres. In addition, a standard network could not capture the fact that the average centre-to-centre spacing of smaller voids should be much less than the corresponding spacing for larger voids, otherwise adjacent small voids would be unreasonably far apart (suggesting an exceptionally low local porosity) and two adjacent large voids would be impossibly close together (suggesting overlap of the two voids).

In order to avoid these problems, a new approach was introduced in \cite{AthWheGra16,Ath17}, where each sphere within the network represents a scaled number of voids, with this scaling number increasing as the sphere size reduces.
The scaling number is
\begin{equation}\label{eq:scalingNumber}
N = \left(\dfrac{r_{\rm sm}}{r_{\rm s}}\right)^3
\end{equation}
where $r_{\rm s}$ is the radius of the sphere and $r_{\rm sm}$ is the mean radius of the spheres.
According to (\ref{eq:scalingNumber}), each sphere of radius smaller than the mean represents more than a single void, whereas each sphere of radius greater than the mean represents less than one void.
This scaled number of voids means that each sphere represents a volume $V_{\rm v}$ of voids given by
\begin{equation}
V_{\rm v} = \dfrac{4}{3} \pi r_{\rm sm}^3
\end{equation}  
Thus, each sphere represents the same volume of voids, whatever its radius. The volume of pipes is ignored in calculating the porosity or degree of saturation of the cell and therefore the porosity $n$ of a cubical cell of side length $l_{\rm CELL}$ containing a total of $m$ spheres is given by
\begin{equation}
n=  \dfrac{4 \pi}{3} \dfrac{m r_{\rm sm}^3}{l_{\rm CELL}^3} 
\end{equation}
The porosity, given by \cite{}, can be set to any desired valye by adjusting the value of $d_{\rm min}$/$r_{\rm sm}$.
Also, the degree of saturation, given by (\ref{eq:saturation}), now simplifies to
\begin{equation} \label{eq:scalingSaturation}
S_{\rm r} = \dfrac{m_{\rm w}}{m}
\end{equation}

The advantage of the void volume scaling approach is that the network model is able to represent a cell containing a large number of small voids without excessive computational effort. In addition, local values of porosity internally within the cell are realistic, rather than being unreasonably low in regions around smaller spheres and impossibly high in regions around the largest spheres. A disadvantage of the void volume scaling approach is, however, that modelling of connectivity between voids, which is always imperfectly represented in a network model, may be represented less realistically than without scaling. This can be illustrated by considering the case of a sphere that is significantly smaller than the mean sphere radius $r_{\rm sm}$. This sphere represents a substantial number $N$ of voids, rather than a single void. The scaling technique means that local connectivity between these $N$ voids is overstated, as they are all assumed to be at the same location in the network and are therefore perfectly connected to each other. In contrast, external connectivity between these $N$ voids and other voids is understated, as only 4 pipes connect the single sphere representing all these $N$ voids to other spheres. Reverse arguments apply to voids that are larger than the mean radius.

\subsection{Conductivity model} \label{sec:transElem}
The water conductivity of the computational cell $k_{\rm CELL}$ is determined by applying a pressure gradient to the periodic cell, which is independent of the uniform capillary suction that is used to control the dying and wetting process of the cell.
The cell conductivity $k_{\rm CELL}$ was evaluated as
\begin{equation}
  k_{\rm {CELL}} = \dfrac{Q_{\rm {CELL}}}{l_{\rm {CELL}} \Delta p_{\rm f}}
\label{eq:macroscopicCond}
\end{equation}
where $Q_{\rm {CELL}}$ is the fluid mass flow rate through a cell face perpendicular to the pressure gradient and $l_{\rm {CELL}}$ is the cell edge length.

The transport through individual pipes is determined by a mass balance equation.
The discrete equation for a pipe is
\begin{equation} \label{eq:elementCon}
\boldsymbol{\alpha}_{\rm e} \mathbf{P} = \mathbf{f}_{\rm e}
\end{equation}
where $\mathbf{P}$ is the vector of water pressures at the nodes of the pipe, $\mathbf{f}_{\rm e}$ is the vector of the nodal flow rate and $\boldsymbol{\alpha}_{\rm e}$ is the conductivity matrix.
The pressures $\mathbf{P}$ are the result of the numerically applied pressure gradient which is used to determine the conductivity of the cell.
The conductivity matrix is evaluated as
\begin{equation}\label{eq:kMatrix}
  \boldsymbol{\alpha}_{\rm e} = \dfrac{\pi r^2_{\rm p}}{h_{\rm p}} k \begin{bmatrix} 1 & -1\\ -1 & 1 \end{bmatrix}
\end{equation}
where $k$ is the conductivity, $r_{\rm p}$ is the pipe radius and $h_{\rm p}$ is the length of the pipe.
The conductivity $k$ has two possible values depending on the type of fluid present in the pipe.
If the pipe is filled with wetting fluid, i.e. the spheres at both ends of the pipe are filled with wetting fluid, the conductivity is equal to the wetting fluid conductivity $k_{\rm l}$. On the other hand, if the pipe is filled with a non-wetting fluid, i.e. at least one of the two spheres is not filled with the wetting fluid, the conductivity is equal to the vapour conductivity $k_{\rm v}$.
For the present study, the wetting fluid is considered to be water and the non-wetting fluid to be air containing water vapour.
The conductivity of a water filled pipe $k_{\rm l}$ is based on Hagen-Poisseuille equation \citep{YioStuBou05}
\begin{equation} \label{eq:kWater}
k_{\rm l} = \dfrac{\rho}{\mu} \dfrac{r_{\rm p}^2}{8}
\end{equation}
where $\rho$ is the density of the water and $\mu$ the dynamic viscosity.
The conductivity of the vapour filled pipe $k_{\rm v}$ is 
\begin{equation}\label{eq:kVapour}
k_{\rm v} = \dfrac{\rho_0}{\rho} \dfrac{D_{\rm v}  \rm{M}}{\rm{R}T}
\end{equation}
where $\rho_0$ is the density of the saturated vapour, $D_{\rm v}$ the molecular diffusivity for vapour through air, M the molar mass of the vapour, R the universal gas constant and $T$ the temperature \citep{EdlAnd43}.
In deriving (\ref{eq:kVapour}), transport due to vapour diffusion in a vapour filled pipe was approximated by a conductivity and flow driven by a water pressure gradient by considering Kelvin's law, which states that under equilibrium conditions across a liquid-vapour interface, the relative humidity of the vapour phase (i.e. the vapour concentration) is related to the capillary suction (i.e. to the water pressure, since air is assumed at uniform pressure).

\section{Analyses and Results}
The influence of cell size $l_{\rm CELL}$ on $S_{\rm r}$ and $k_{\rm {CELL}}$ is investigated for varying capillary suction $P_{\rm c}$ representing drying and wetting processes.
For this, three cell sizes with ratios $l_{\rm {CELL}}/d_{\rm{min}} = $~3, 5 and 10 were chosen, and 100 cells (each with a different set of sphere locations and different sets of sphere and pipe radii) were generated for each cell size so that the COVs of $S_{\rm r}$ and $k_{\rm CELL}$ were determined accurately.
The analysis of each cell began from an almost saturated condition at $S_{\rm r} \approx 0.99$ (with only seeding cells filled with drying fluid, i.e. air).
Then, the cell was fully dried (to $S_{\rm r} \approx 0.01$) and fully wetted again (to $S_{\rm r} \approx 0.99$) by changing $P_{\rm c}$.
For each value of $P_{\rm c}$, the values of $S_{\rm r}$ and $k_{\rm {CELL}}$ were evaluated for each cell.
Then a mean and a COV of each quantity was evaluated.
The solution procedure is described in detail in the Appendix~\ref{sec:implementation}.

The input parameters used in the analyses consist of geometrical parameters related to the size distributions of spheres and pipes (Table~\ref{tab:inputNetwork}), and input parameters required related to the fluids and their interaction with the void network (Table~\ref{tab:inputPhysical}). For water, the parameters are the density $\rho$, the absolute (dynamic) viscosity $\mu$, the contact angle $\theta$ and the surface tension $\gamma$.
For the vapour, the parameters are the saturated vapour density $\rho$, the water vapour diffusivity $D_{\rm v}$, temperature $T$, universal gas constant~R and the molar mass~M. The parameters in Table~\ref{tab:inputPhysical} have a strong physical meaning with input values that can be found in the literature.
\begin{table}
\caption{Network input parameters}
\label{tab:inputNetwork}
\centering
\begin{tabular}{lll}
\\
Network parameter & value & units\\\hline
Minimum distance of Delaunay points $d_{\rm {min}}$ &  $5.6 \times 10^{-4}$ & m\\
Mean of pipe radii distribution  $r_{\rm pm}$ & $10^{-6}$  & m\\
COV of pipe radii distribution  COV$^{\rm p}$ & 1 & \\
Minimum pipe radius $r_{\rm p}^{\rm min}$ & $1.35 \times 10^{-9}$ & m\\
Maximum pipe radius $r_{\rm p}^{\rm max}$& $2 \times 10^{-3}$ &m\\
Mean of sphere radii distribution $r_{\rm sm}$ & $10^{-5}$ & m\\
COV of sphere radii distribution COV$^{\rm s}$ & 1 & \\
Minimum sphere radius $r_{\rm s}^{\rm min}$& $1.35 \times 10^{-9}$&m\\
Maximum sphere radius $r_{\rm s}^{\rm max}$& $2 \times 10^{-3}$&m\\
\end{tabular}
\end{table}
\begin{table}
\caption{Standard physical input parameters.}
\label{tab:inputPhysical}
\centering
\begin{tabular}{llll}
\\
Physical parameters & Value & Units & Reference\\\hline
Liquid water density $\rho$ & 1000 & kg/m$^3$ & \citet{DauFra77}\\
Saturated gas fluid density  $\rho_{0}$ & 0.01724 & kg/m$^3$ & \citet{MayRog76}\\
Dynamic viscosity $\mu$ & 0.001002 & Pa s & \citet{DauFra77}\\
Temperature $T$&293&K& \citet{DauFra77}\\
Gas universal constant~R&8.314&J/mol/K & \citet{DauFra77}\\
molar mass of vapour~M&0.018&kg/mol & \citet{DauFra77}\\
Surface tension $\gamma$ & 0.072 & N/m & \citet{DauFra77}\\
Water molecular diffusivity $D_{\rm v}$&$2.3\times10^{-5}$& $\rm {m^2}/\rm{s}$ & \citet{Hag03} \\
Contact angle $\theta$ & 0 & $^{\circ}$ &\citet{Del15}\\
\end{tabular}
\end{table}

\begin{figure}
\centerline{
\includegraphics[width=12cm]{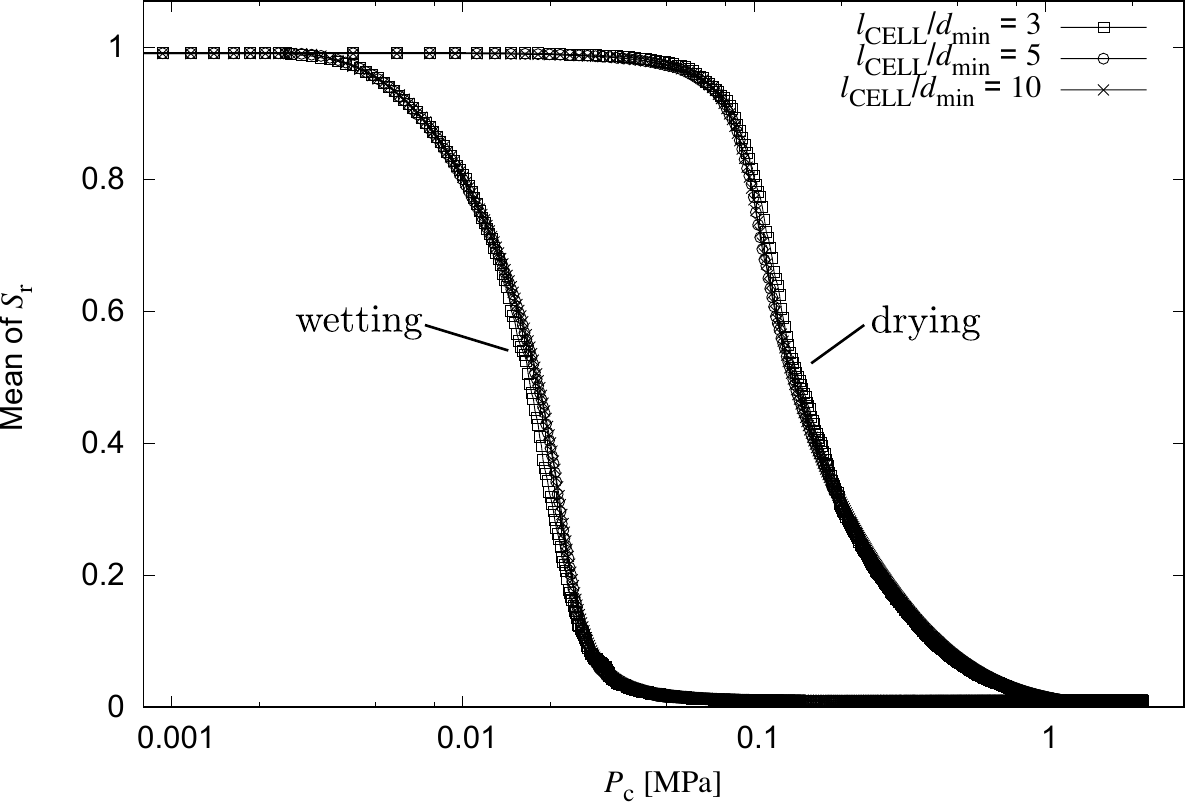}
}
\caption{Influence of cell size on retention during drying and wetting: mean degree of saturation $S_{\rm r}$ versus capillary suction $P_{\rm c}$ for 100 realisations for $l_{\rm CELL}/d_{\rm {min}} = $~3, 5 and 10.}
\label{fig:meanSaturation}
\end{figure}

\begin{figure}
\centerline{
\includegraphics[width=12cm]{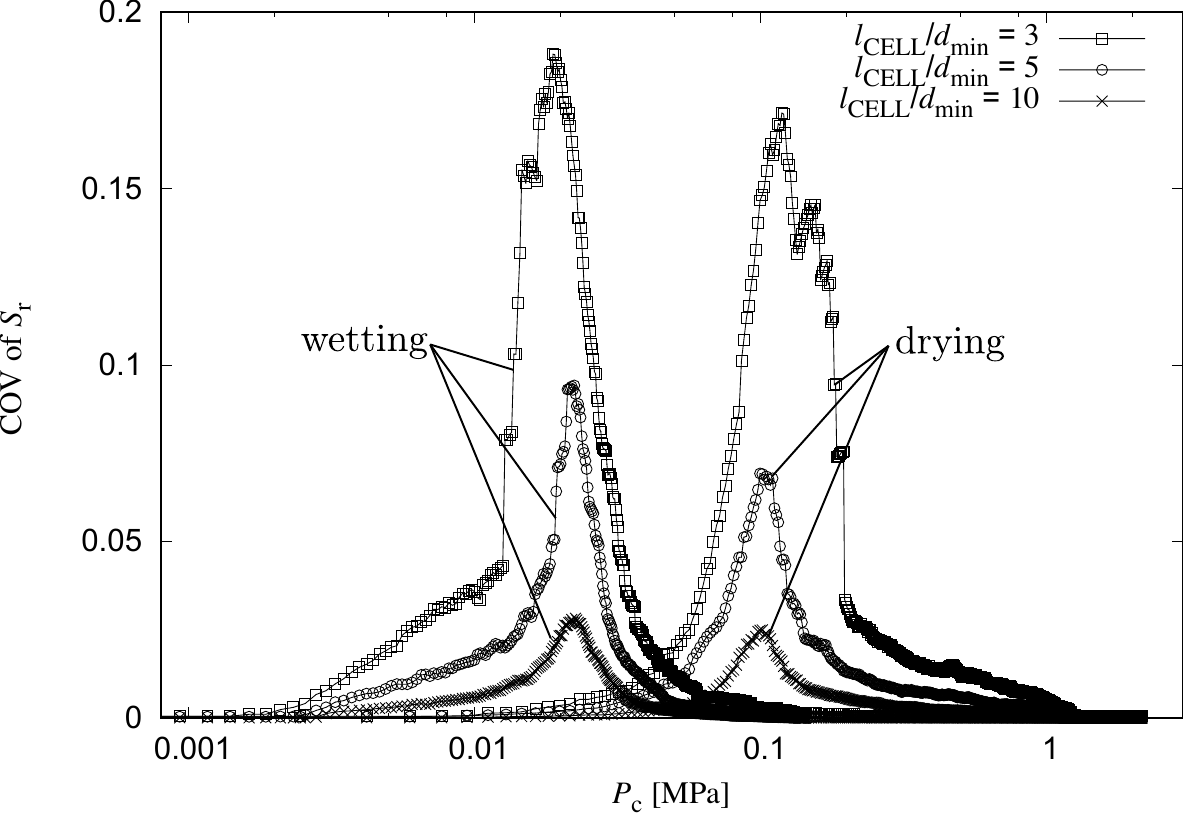}
}
\caption{Influence of cell size on retention during drying and wetting: coefficient of variation of degree of saturation (COV of $S_{\rm r}$) versus capillary suction $P_{\rm c}$ for 100 realisations for $l_{\rm CELL}/d_{\rm {min}} =$~3,~5~and~10.}
\label{fig:COVSaturation}
\end{figure}

In Figure~\ref{fig:meanSaturation}, means of $S_{\rm r}$ during the drying and wetting are plotted against $P_{\rm c}$ for the three cell sizes.
The curves of the means of $S_{\rm r}$ for the three cell sizes are almost indistinguishable.
Hence, for $l_{\rm CELL}/d_{\rm min} \geq 3$ there is a small influence of cell size on the mean values of $S_{\rm r}$ during drying.
There is a strong influence of the cell size on the COV of $S_{\rm r}$ during drying and wetting for intermediate $S_{\rm r}$ values as shown in Figure~\ref{fig:COVSaturation}.
For increasing cell size, the COV decreases.
The increase and then decrease of COV of $S_{\rm r}$ with $P_{\rm c}$ increase is attributed to the change of the negative derivative of degree of saturation with respect to capillary suction (in the semi-logarithmic plot shown in Figure~\ref{fig:meanSaturation}).
The COV of $S_{\rm r}$ increases with an increase of this gradient.
The peak of the COV curve exactly coincides with the point of maximum gradient on the retention curve (in a semi-logarithmic plot). 
The fact that the mean is almost independent of the cell size and the COV decreases with increasing cell size indicates that an RVE exists for water retention during drying and wetting. 

\begin{figure}
\centerline{
\includegraphics[width=12cm]{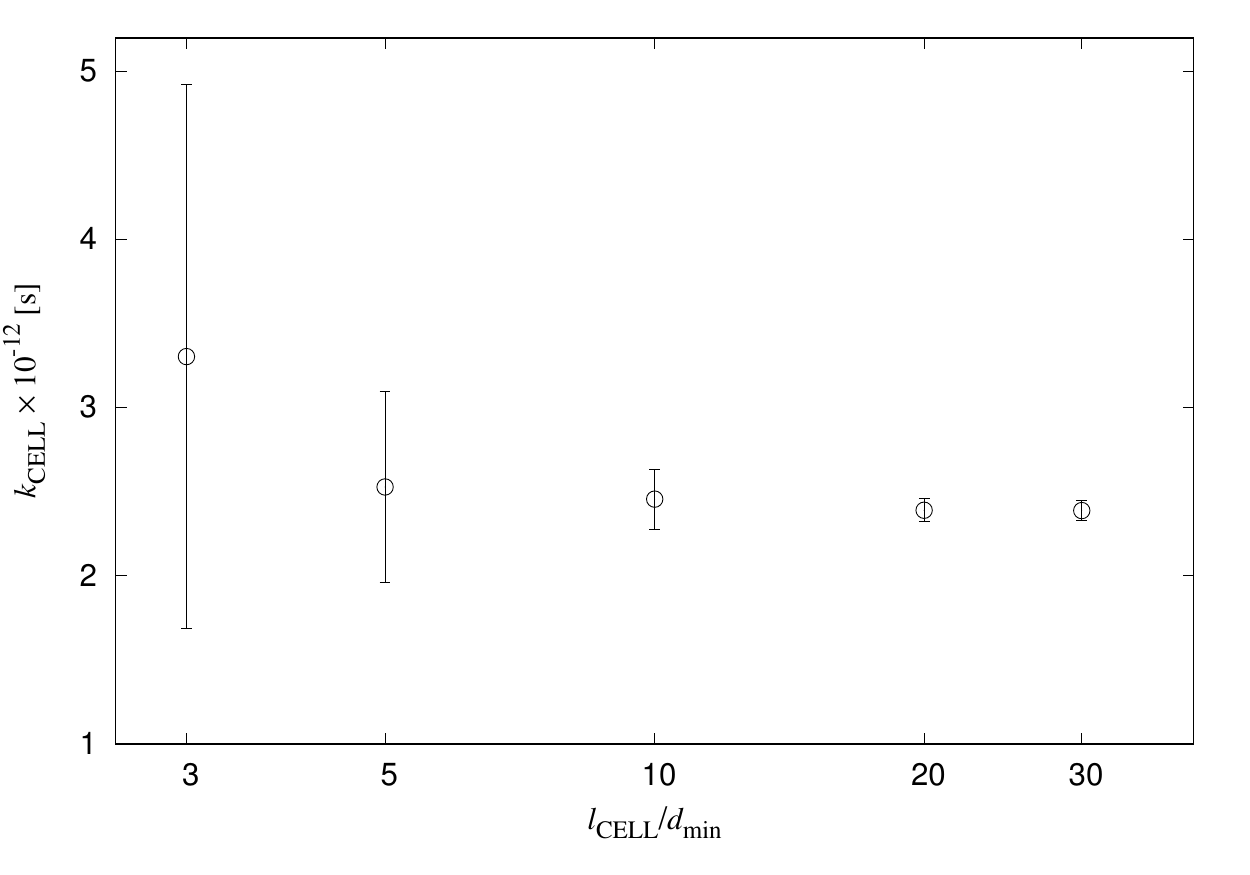}
}
\caption{Influence of cell size on conductivity $k_{\rm{CELL}}$ for $S_{\rm r} \approx$ 0.99: the symbols represent the mean of 100 realisations and the error bars show $\pm$ one standard deviation.}
\label{fig:condSat}
\end{figure}

The influence of cell size on conductivity was carried out in two steps.
Firstly, the influence of cell size on the mean and standard deviation of the conductivity for a constant degree of saturation of $S_{\rm r} \approx$ 0.99 (i.e. almost saturated conditions) was investigated for a large range of $l_{\rm{CELL}}/d_{\rm{min}}$.
Then, the conductivity curves for unsaturated conditions for drying and wetting were determined for the three cell sizes used for the retention behaviour.
For $S_{\rm r} \approx$ 0.99, the conductivity $k_{\rm{CELL}}$ was determined for $l_{\rm {CELL}}$/$d_{\rm {min}} =$~3, 5, 10, 20 and 30.
For each cell size, 100 different realisations were analysed (each with a different set of sphere locations and different sets of sphere and pipe radii).
Figure~\ref{fig:condSat} presents the variation of the mean of $k_{\rm {CELL}}$ with the normalised cell size $l_{\rm {CELL}}/d_{\rm {min}}$.
The error bars show $\pm$ one standard deviation.
The standard deviation decreases with increasing cell size and the mean values decreases with increasing cell size for $3 < l_{\rm {CELL}}/d_{\rm {min}} < 10$ and then remains almost constant for $l_{\rm {CELL}}/d_{\rm {min}}>10$.
Thus, for this almost saturated condition, the COV decreases with increasing cell size as well.
The decreasing COV with increasing cell size is an expected behaviour, because increasing size means that more statistical information is included and therefore there is less difference in the overall behaviour from cell to cell.
The results confirm findings in the literature \citep{ZhaZhaChe00,DuOst06} that there is a RVE for conductivity for saturated conditions.

\begin{figure}
\centerline{
  \includegraphics[width=12cm]{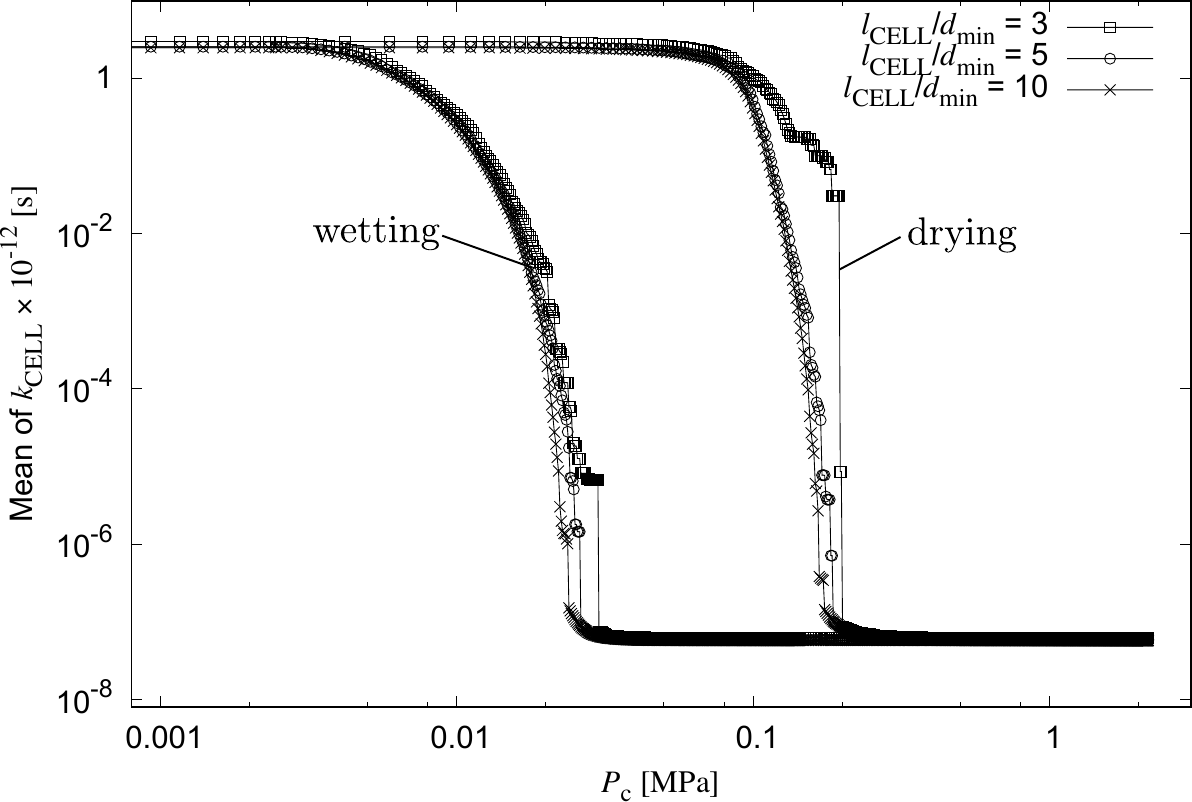}
}
\caption{Influence of cell size on mean conductivity during drying and wetting: conductivity $k_{\rm {CELL}}$ (in log scale) versus capillary $P_{\rm c}$ for 100 realisations for $l_{\rm CELL}/d_{\rm {min}} =$~3, 5 and 10.}
\label{fig:MeanConductivityLog}
\end{figure}

For unsaturated conditions, the mean and COV of $k_{\rm {CELL}}$ were evaluated for $l_{\rm {CELL}}$/$d_{\rm {min}} =$~3, 5, and 10 as part of the same analyses as those presented for retention.
For each value of $P_{\rm c}$ and each cell size the mean and COV of $k_{\rm {CELL}}$ were evaluated for 100 realisations.
The mean value of $k_{\rm {CELL}}$ is plotted against $P_{\rm c}$, for drying (increasing capillary suction) and wetting (decreasing capillary suction) in Figure~\ref{fig:MeanConductivityLog}.
As the cell size increases, the conductivity curves come closer.
For $P_{\rm c}$ values that correspond to almost saturated conditions, i.e. for low suction values, and for those that correspond to almost dry conditions, i.e. for high suction values, the mean $k_{\rm {CELL}}$ decreases with increasing cell size as expected from the previous saturated analysis (Figure~{\ref{fig:condSat}}).
However, between these two stages there exist a narrow range of suction for which there are differences of orders of magnitude between the mean conductivities for the same $P_{\rm c}$ for the three cell sizes.
These high differences are attributed to the effect of onset of percolation which is the stage at which a continuous path of wetting fluid that connects opposite faces (in the direction of the gradient applied) is first formed during wetting (or finally lost during drying). In each individual realisation, a dramatic change of $k_{\rm {CELL}}$ occurs when the continuous path of wetting fluid is first formed (during wetting) or finally lost (during drying), but this can occur at different values of $P_{\rm c}$ in different realisations, and also, on average, it will occur at higher values of $P_{\rm c}$ at smaller cell sizes.

\begin{figure} [!ht]
\centerline{
\begin{tabular}{cc}
  \includegraphics[width=12cm]{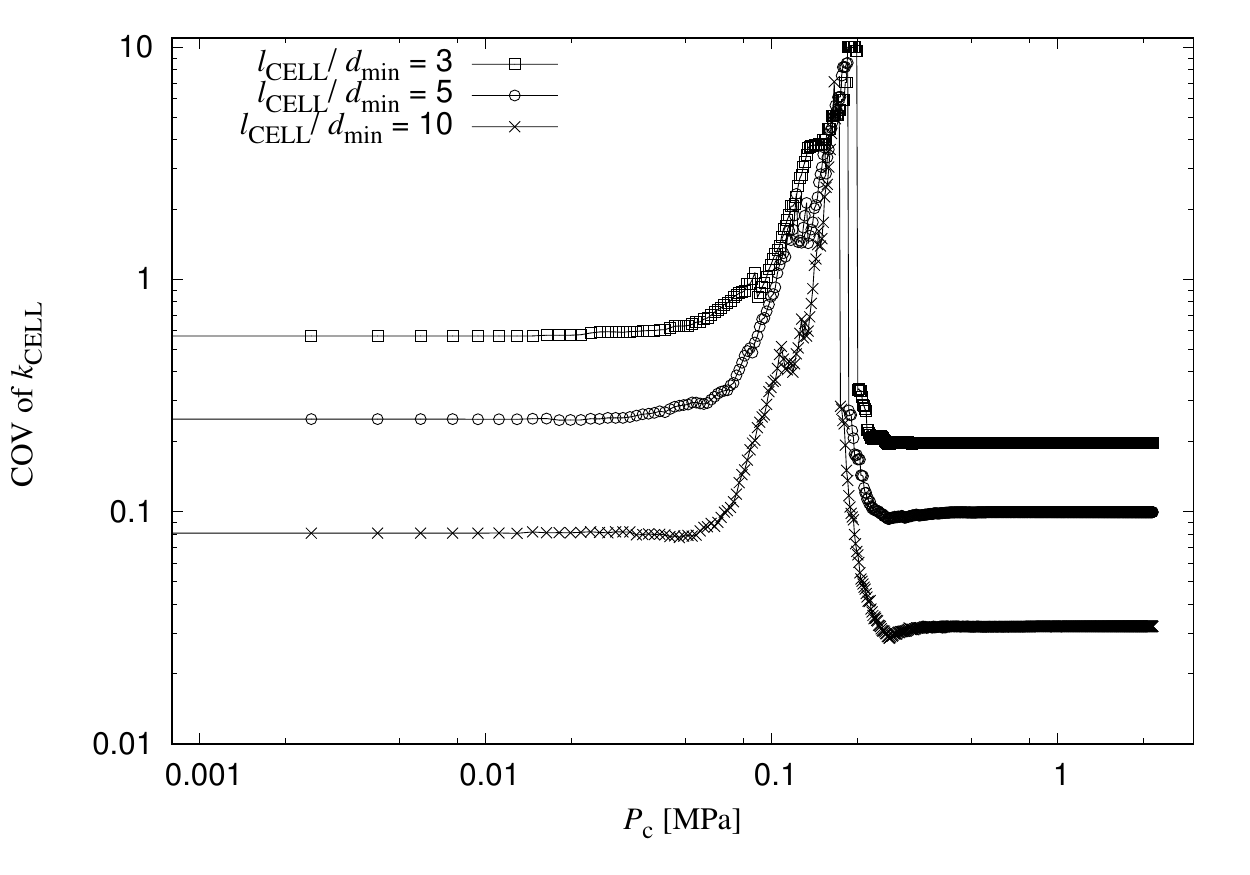}\\
  (a)\\
  \includegraphics[width=12cm]{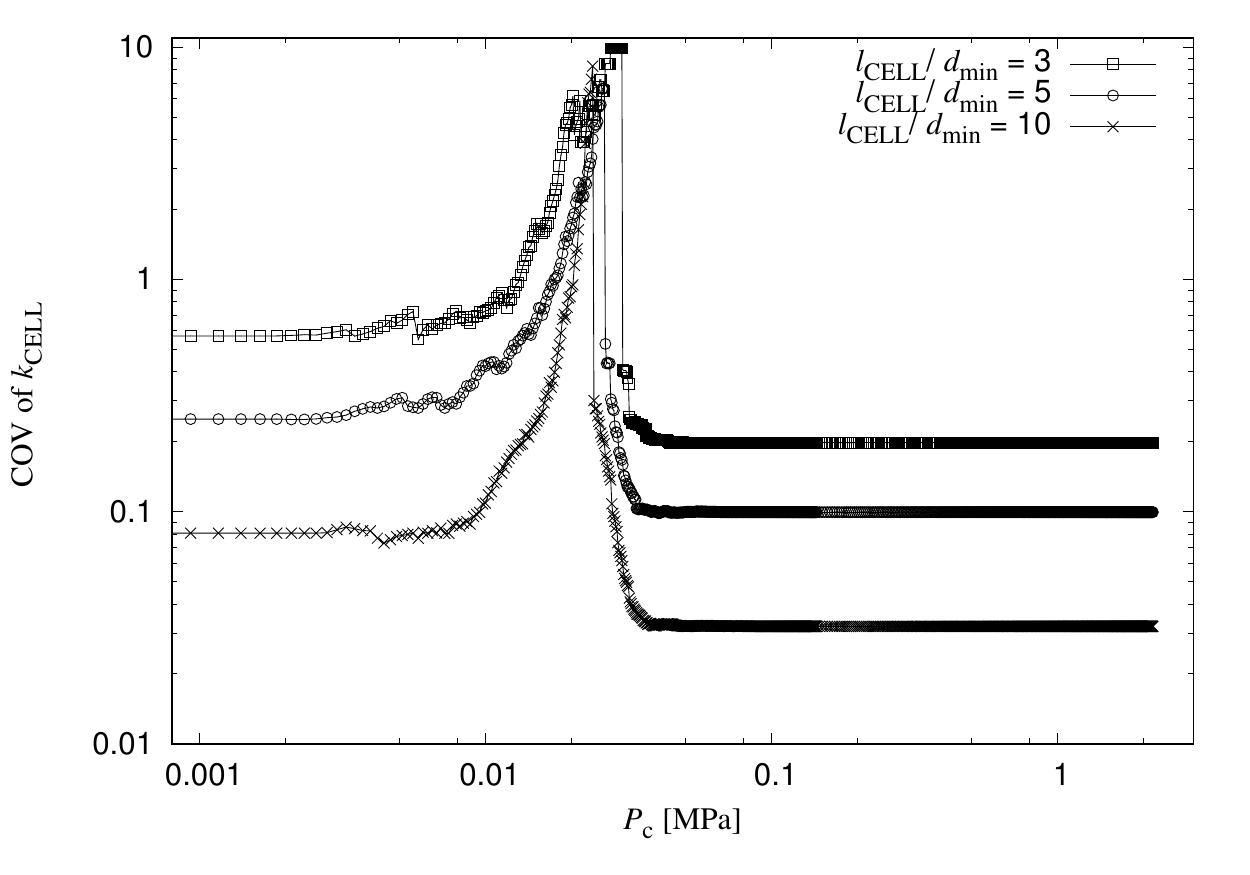}\\
  (b)
\end{tabular}
}
\caption{Influence of cell size on conductivity during (a) drying and (b) wetting: COV of $k_{\rm CELL}$ versus capillary suction $P_{\rm c}$ for 100 realisations for $l_{\rm CELL}/d_{\rm {min}}=$~3, 5 and 10.}
\label{fig:COVCond}
\end{figure}

The COV of $k_{\rm{CELL}}$ versus $P_{\rm c}$ is presented in Figure~\ref{fig:COVCond}.
Generally, the COV of $k_{\rm {CELL}}$ for most values of $P_{\rm c}$ decreases with increasing cell size, as expected.
Furthermore, it can be observed that for almost saturated conditions (low values of $P_{\rm c}$), the COV of $k_{\rm {CELL}}$ is larger than that for almost dry conditions (high values of $P_{\rm c}$), because the flow rate along a single pipe in dry conditions is determined by the square of the pipe radius whereas for the saturated case it is determined by the fourth power of the pipe radius (see (\ref{eq:kMatrix}), (\ref{eq:kWater}) and (\ref{eq:kVapour})).

Both for drying and wetting curves in Figure~\ref{fig:COVCond}, there is a narrow range of $P_{\rm c}$ over which the COV of $k_{\rm CELL}$ is very large and does not decrease with increasing cell size.
Over this narrow range of $P_{\rm c}$ there is no size that satisfies the conditions of an RVE used for this study.
This lack of RVE is due to the onset of percolation.
Outside this narrow range of $P_{\rm c}$, the COV of $k_{\rm CELL}$ is low and decreases with increasing cell size.

\section{Conclusions}
Random network transport analyses of unsaturated porous materials with different cell sizes resulted in the following conclusions:
\begin{itemize}
\item There is a representative volume element for water retention, since, for any value of capillary suction, the coefficient of variation of degree of saturation decreases and the mean of degree of saturation approaches a constant value with increasing cell size. 
\item  For conductivity, there is a range of capillary suctions at which the COV of conductivity does not decrease with increasing cell size. This narrow capillary suction range corresponds to the stage at which onset of percolation occurs. Outside this range, the mean of conductivity for any value of capillary suction approaches a constant value and the COV of conductivity decreases with increasing cell size, which indicates that an RVE exists.
\end{itemize}

\subsection*{Acknowledgement}
The numerical analyses were performed with the nonlinear analyses program OOFEM \cite{Pat12} extended by the present authors to discrete network modelling. 
This extended version of OOFEM is available at http://doi.org/10.5281/zenodo.1222831.
The authors acknowledge funding received from the UK Engineering and Physical Sciences Research Council (EPSRC) under grant EP/I036427/1 and funding from Radioactive Waste Management Limited (RWM) (http://www.nda.gov.uk/rwm), a wholly-owned subsidiary of the Nuclear Decommissioning Authority. RWM is committed to the open publication of such work in peer reviewed literature, and welcomes e-feedback to rwmdfeedback@nda.gov.uk.

\bibliographystyle{elsarticle-harv}
\bibliography{general}

\section*{Appendix: Solution procedure}\label{sec:implementation}
The transport problem is solved as a stationary problem.
At each step $h+1$ of an analysis, the driving variable is the target capillary suction $P_{\rm c}^{{\rm{T}}(h+1)}$.
For each $P_{\rm c}^{{\rm{T}}(h+1)}$, the configuration of the drying and wetting fluids is determined.
Then, the degree of saturation $S_{\rm r}$ is evaluated according to (\ref{eq:scalingSaturation}) presented in Section~\ref{sec:retentionModel}.
For the conductivity evaluation, the configuration of the fluids is fixed and the network is subjected to a notional uniaxial gradient of capillary suction.
The resulting network flow rate is computed as the reaction total flow in the direction in which the uniaxial gradient was applied.

Since periodic boundary conditions are used, a special treatment of the network initiation is required , which is based on seeding spheres as described in Section~\ref{sec:model}.
For simulation of a wetting process from initially ``dry'' conditions, the wetting process must be preceded by a drying process, as described previously.
Therefore, network begins with all its spheres filled with wetting fluid except for one.
Spheres are then filled with drying fluid one by one.
The sphere chosen to fill with drying fluid each time, amongst all the spheres in the network, is the one connected to a drying fluid filled neighbour that has the lowest value of emptying capillary suction according to (\ref{eq:LY}), where $r$ in (\ref{eq:LY}) is taken as the radius of the connecting pipe.
Once a number $m_{\rm {seed}}$ of spheres remain filled with wetting fluid, the seeding process is completed.
The remaining spheres still filled with wetting fluid are the seeding spheres for the subsequent wetting process.

During the wetting process, the sphere chosen to fill next with wetting fluid is the one connected to a wetting fluid filled neighbour that has the highest value of filling capillary suction according to (\ref{eq:LY}) (where $r$ in (\ref{eq:LY}) is taken as the radius of the sphere about to be filled).
The wetting process is terminated when a number $m_{\rm {seed}}$ of spheres are left filled with drying fluid.
These would be the seeding spheres for simulation of any subsequent drying process from a ``fully wet'' condition.

Five quantities are assigned to each sphere.
\begin{enumerate}
\item A flag that states whether the sphere is filled with wetting fluid or drying fluid, i.e. 0 when filled with wetting fluid and 1 when filled with drying fluid. 
\item The total number of neighbouring spheres $N_{\rm {n}}$.
\item The number of these neighbours that are filled with the wetting fluid $N_{\rm {wn}}$.
\item The capillary suction value ($P_{\rm cD}$) that the sphere will fill with drying fluid, determined by the radius of the largest pipe connecting it to a neighbouring sphere filled with drying fluid. 
\item The capillary suction value ($P_{\rm cW}$) that the sphere will fill with wetting fluid, if it is connected to a neighbouring sphere filled with wetting fluid, determined by the radius of the sphere under consideration.
\end{enumerate}

In the beginning of the analysis, the above quantities are assigned as described below.
All spheres are considered to be filled with the wetting fluid, hence a flag value equal to 0 is assigned for each of them.
At all spheres, the values of $N_{\rm {n}}$ and $P_{\rm cW}$ are assigned.
Additionally, a very large value for $P_{\rm cD}$ is initially assigned at each sphere. 
This is done because a real value of $P_{\rm cD}$ cannot yet be determined for any of the spheres, since none of them are yet available to be filled with drying fluid (none of them have neighbouring spheres filled with drying fluid).
For each sphere the equality $N_{\rm {wn}}=N_{\rm {n}}$ is set. 
Both these values cannot be less than zero.
Throughout the analysis and the initiation process the values of $N_{\rm {n}}$ and $P_{\rm cW}$ of each sphere are invariant.
The tracking of the pipe filling is not needed since it does not play an explicit role in determining the process of emptying or filling of spheres.

\begin{figure}[!ht]
\begin{center}
\includegraphics[height=8.cm]{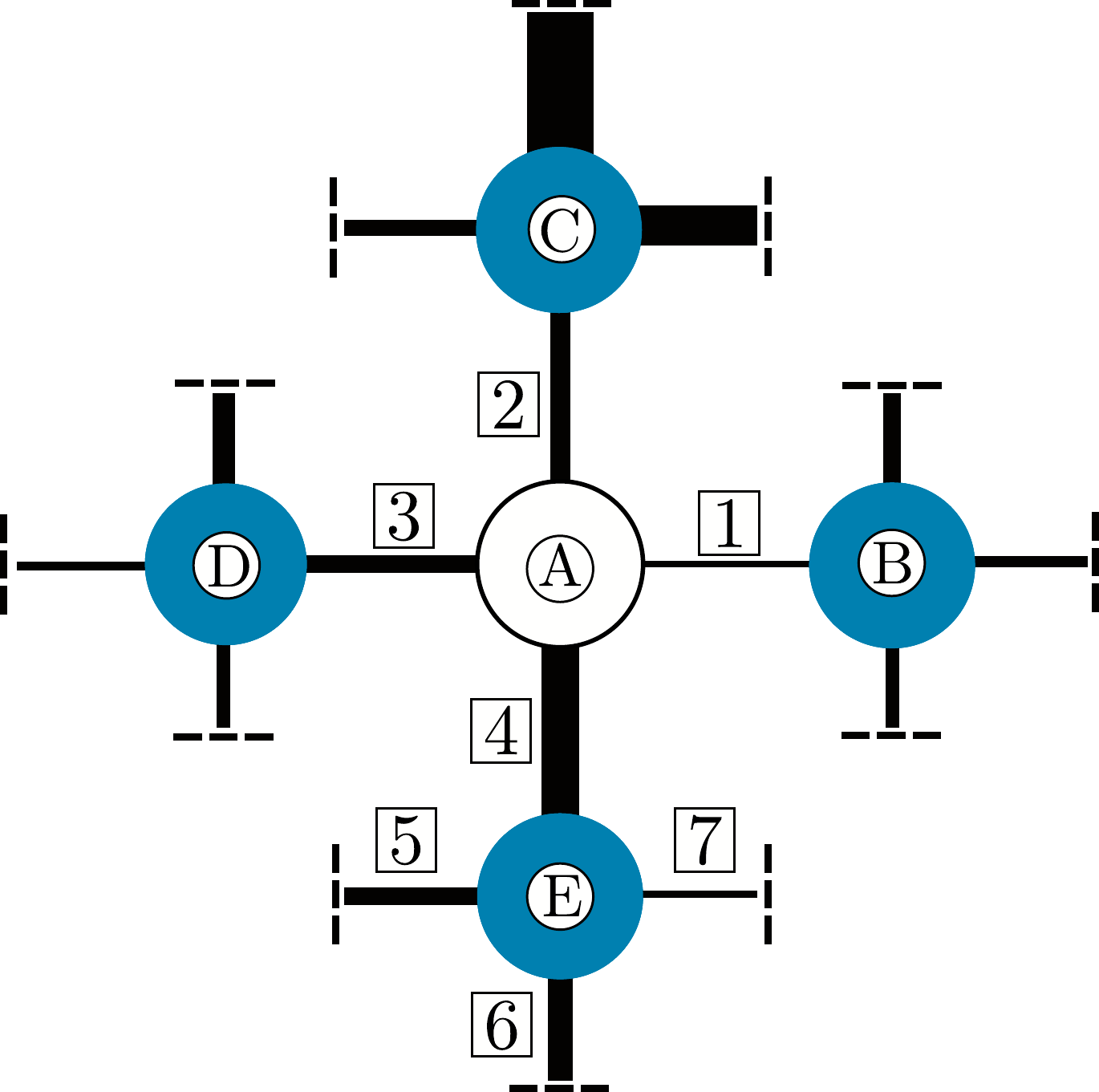} \\
\caption{Schematic presentation of a sub-pore network: Drying fluid filled sphere A, wetting fluid filled spheres B, C, D and E, and numbered pipes.}
\label{fig:one_sphere_connected}
\end{center}
\end{figure}

After setting all the aforementioned initial values, the seeding process begins.
A single sphere filled with drying fluid serves the purpose of the first seeding sphere, and a flag equal to 1 is assigned to this sphere.
The neighbours of the first seeding sphere are identified and the value of their $N_{\rm {wn}}$ is decreased by one.
Figure~\ref{fig:one_sphere_connected} presents a schematic 2-D representation of the initial seeding sphere A and its neighbours B, C, D and E connected through pipes 1, 2, 3 and 4, respectively.  
All spheres in the network have a coordination number $N_{\rm n}$ equal to four.
The blue and white spheres represent the wetting and drying fluid filled ones, respectively.
The $P_{\rm cD}$ values of B, C, D and E are changed from their initial very large value to that corresponding to pipes that connect them to sphere A.

\begin{figure}[!ht]
\begin{center}
\includegraphics[height=8.cm]{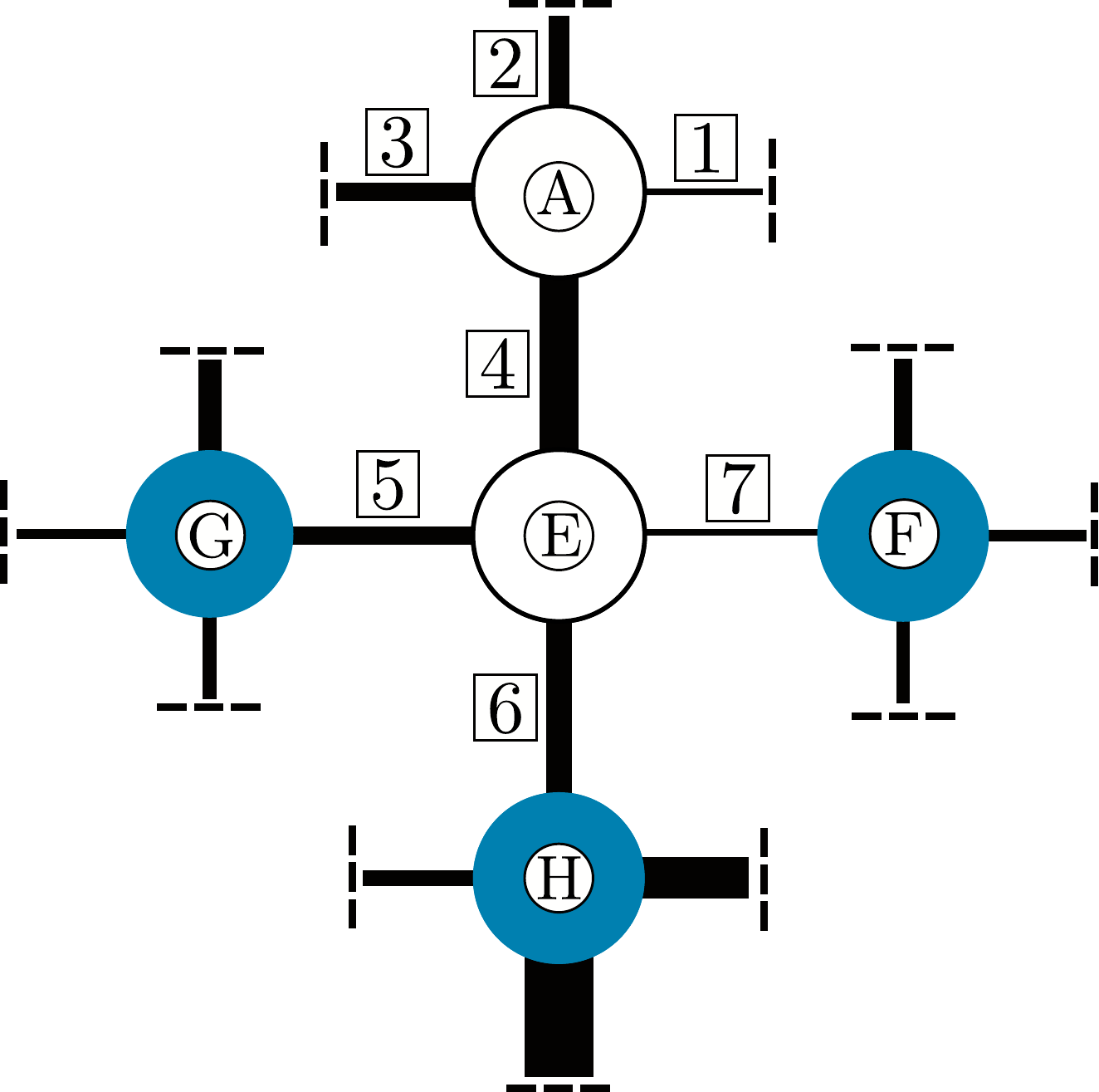} \\
\caption{Schematic presentation of a sub-pore network: Drying fluid filled spheres A and E, wetting fluid filled spheres G, F and H, and numbered pipes.}
\label{fig:two_sphere_connected}
\end{center}
\end{figure}

All spheres with $N_{\rm {wn}} < N_{\rm {n}}$ and their phase condition flag equal to 0 are now candidates to fill with drying fluid at the next step of the initiation process.
Hence, spheres B, C, D, and E in Figure \ref{fig:one_sphere_connected} are the only candidates at the first step of the initiation.
The sphere with the lowest $P_{\rm cD}$ value fills with drying fluid next.
Hence, sphere E fills with drying fluid since pipe 4 has the largest radius amongst all the other connected ones.
The flag value for sphere E therefore switches to 1.

Figure \ref{fig:two_sphere_connected} shows sphere E, at the second step of the analysis, with its four neighbour spheres A, F, G and H.
The $N_{\rm {wn}}$ values of each neighbour of sphere E are decreased by one.
New paths for drying fluid to flow and potentially fill the neighbours of sphere E are now available.
Therefore, the values of $P_{\rm cD}$ already assigned to spheres A, F, G and H are compared to the drying capillary suction value due to the new available pathways through pipes 4, 7, 5 and 6, respectively. 
If the capillary suction value corresponding to the new pathway is lower than the previously assigned one, then $P_{\rm cD}$ is assigned this new value.

This process is continued until the entire cell is considered to be dry i.e. until the number of wetting fluid filled spheres are equal to the selected seeding number $m_{\rm {seed}}$.
This is done in a number ($m$~-~$m_{\rm {seed}}$) sphere emptying events, each involving filling of an individual sphere with drying fluid.
At each event, the sphere with the lowest value of $P_{\rm {cD}}$ amongst the candidate spheres is the one chosen to fill with drying fluid. 

Once the drying process is completed, in the case that the user chooses to start an analysis from wet conditions, the cell is then subjected to wetting.
During wetting, a similar process to the one previously described is performed.
At each of a total of $m \rm{-2} m_{\rm{seed}}$ sphere filling events all the spheres of the network are checked to see whether they are candidates to be the next one for filling with wetting fluid.
This check requires the sphere to be currently filled with drying fluid ($\rm{flag = 1}$) and for it to be connected to a wetting fluid filled neighbour ($N_{\rm {wn}}>0$).
For each individual sphere filling event, the sphere that fulfils the aforementioned criteria and has the highest $P_{\rm cW}$ value is the one that fills with wetting fluid.

At each step of the analysis, the user is able to determine the increment of capillary suction of the transport network.
The value of capillary suction resulting from the chosen increment at an arbitrary step $(h+1)$ of the analysis is denoted as $P_{\rm c}^{{\rm T}(h+1)}$ which will be referred from now on as target capillary suction.
For a fluid configuration resulting from the previous step $(h)$, the wetting fluid filled sphere with the lowest drying capillary suction $P_{\rm {cD}}$ value and connected to a drying fluid filled one, determines the value ${P_{\rm {cD}}}_{\rm min}$.
Also, at the same step, the drying fluid filled sphere with the highest wetting capillary suction value $P_{\rm {cW}}$ and connected to a wetting fluid filled sphere determines the value max ${P_{\rm {cW}}}_{\rm max}$.
For the configuration to be stable, the relation below should hold
\begin{equation}
 {P_{\rm {cW}}}_{\rm max} < P_{\rm c}^{{\rm T}(h+1)} < {P_{\rm {cD}}}_{\rm min}
\label{fm:Stable}
\end{equation}
When (\ref{fm:Stable}) holds, no filling of spheres with drying or wetting fluid occurs.
However, when (\ref{fm:Stable}) does not hold, a new configuration should be searched until (\ref{fm:Stable}) is fulfilled.
When $P_{\rm c}^{{\rm T}(h+1)} < {P_{\rm {cW}}}_{\rm max}$ the drying fluid filled sphere with $P_{\rm cW} = {P_{\rm {cW}}}_{\rm max}$ is subjected to wetting.
After this sphere filling with wetting fluid, a new ${P_{\rm {cW}}}_{\rm max}$ value is evaluated since the sphere for which previously $P_{\rm cW} = {P_{\rm {cW}}}_{\rm max}$ is now filled with wetting fluid.
Similarly, when $P_{\rm c}^{{\rm T}(h+1)}>{P_{\rm {cD}}}_{\rm min}$ a sphere is subjected to drying and a new ${P_{\rm {cD}}}_{\rm min}$ is found.
After each filling with wetting or drying fluid of a sphere, (\ref{fm:Stable}) is checked once more.
If (\ref{fm:Stable}) is not fulfilled, another sphere is chosen to be filled with wetting fluid for $P_{\rm c}^{{\rm T}(h+1)} < {P_{\rm {cW}}}_{\rm max}$ or drying fluid for  ${P_{\rm {cD}}}_{\rm min} < P_{\rm c}^{{\rm T}(h+1)}$.
Therefore, for a given $P^{{\rm T}(h+1)}$, more than one sphere change can occur.
za
The description of the above process assumes that ${P_{\rm {cW}}}_{\rm max} < {P_{\rm {cD}}}_{\rm min}$ and also that a configuration that fulfils (\ref{fm:Stable}) can be found.
However, due to the random assignment of spheres and pipes, one or both of those statements might not hold at a step $(h+1)$.
For resolving these problems different rules are applied to determine the configuration at step $(h+1)$.

When ${P_{\rm {cD}}}_{\rm min} < {P_{\rm {cW}}}_{\rm max}$ the type of fluid filling of spheres in the network is the same as the last one performed in the previous trial sphere filling.
In the case that the previous trial involved a sphere filling with wetting fluid, the next sphere filling will be with wetting fluid as well.
Reverse arguments apply if the previous trial involved a sphere filling with drying fluid.
If the very first configuration search at step $(h+1)$ involves the condition ${P_{\rm {cD}}}_{\rm min} < {P_{\rm {cW}}}_{\rm max}$, then the choice of the fluid that is intruding is the same as that for the last sphere that was filled in the previous step $(h)$. 

Another issue that might arise is that of having configurations that repeat at the same step $(h+1)$ i.e. the pair of spheres that correspond to ${P_{\rm {cW}}}_{\rm max}$ and ${P_{\rm {cD}}}_{\rm min}$ repeat throughout a configuration search.
In this case the search of the configuration would infinitely continue without fulfilling (\ref{fm:Stable}).
Hence, when a configuration is reached for the second time, the search is stopped, and the capillary suction is changed to the next target value. 

\end{document}